# Entropic formulation for the protein folding process: hydrophobic stability correlates with folding rates

J. P. Dal Molin[*] and A. Caliri

*Departamento de Física e Química, FCFRP, Universidade de São Paulo, 14040-903 Ribeirão Preto, SP, Brazil*

[*] jpdm@usp.br

## ABSTRACT

We assume that the protein folding process follows two autonomous steps: the conformational search for the native, mainly ruled by the hydrophobic effect; and, the final adjustment stage, which eventually gives stability to the native. Our main tool of investigation is a 3D lattice model provided with a ten-letter alphabet, the stereochemical model. This model was conceived for Monte Carlo (MC) simulations when one keeps in mind the kinetic behavior of protein-like chains in solution. In order to characterize the folding characteristic time ($\tau$) by two distinct sampling methods, first we present two sets of $10^3$ MC simulations for a fast protein-like sequence. For these sets of folding times, $\tau$ and $\tau_q$ were obtained with the application of the standard Metropolis algorithm (MA), and a modified algorithm ($M_qA$). The results for $\tau_q$ reveal two things: *i*) the hydrophobic chain-solvent interactions plus a set of inter-residues steric constraints are enough to emulate the first stage of the process: for each one of the $10^3$ MC performed simulations, the native is always found without exception, *ii*) the ratio $\tau_q/\tau \cong 1/3$ suggests that the effect of local thermal fluctuations, encompassed by the Tsallis weight, provides an innate efficiency to the chain escapes from energetic and steric traps. A physical insight is provided. Our second result was obtained through a set of 600 independent MC simulations performed with the $M_qA$ method applied to a set of 200 representative targets (native structures). The results show how structural patterns modulate $\tau_q$, which cover four orders of magnitude in the temporal scale. The third, and last result, was obtained from a special kind of simulation for those same 200 targets, we simulated their stability. We obtained a strong correlation ($R=0.85$) between the hydrophobic component of protein stability and the folding rate: the faster is the protein to find the native, larger is the hydrophobic component of its stability. This final result suggests that the hydrophobic interactions could not be a general stabilizing factor for proteins.

# 1 – Introduction

The protein folding problem has been tackled for decades under different conceptual perspectives and varied experimental techniques. Regardless all the effort, it remains as one of the most important open problems in structural biology [1-3]. Some authors have supported that the hydrophobic effect should be the main driving force that promotes the correct collapse of chain apolar groups, as well as the main responsible for the protein stabilization [4-6]. On the other hand, some authors argue that intrapeptide hydrogen bonds (HB) could be the main driving-force for folding reaction, because it seems to be enthalpic favored over the peptide-water HB by about 1.5 kcal/mol [7-9]. The second hypothesis inspired the proposition of backbone-based theories for protein folding [8, 9]. The persistence of opposite views about fundamental determinants of the folding suggests that both alternatives may have key roles along the process. However, how? Here we explore the assumption that the hydrophobic effect should be the main driving force in early events of the folding process, the search stage for the native structure, when chain configurations populate the unfolded state most of times.

Our basic assumption is that, from a mechanistic point of view, the complete folding process of a small, single-domain globular protein should consists of two independent and consecutive steps [10,11]: (*i*) the conformational *search* mechanism, mainly ruled by hydrophobic interactions between solvent-chain, which should produce a near-native globule, followed by (*ii*) the final adjustment stage in which extra forces should provide effective stability to the protein. In the first stage, hydrophobic interactions and steric specificities are conjectured to be the only efficient agents who guide the collapse of apolar groups into a conformation close to the native, which seems to operate as a global trap. Then, it is assumed that only when close enough of the native conformation all stabilizing factors lead to the final state, in which most of hydrophobic elements of the chain are buried in the globule, backbone HBs are right oriented and protected from the solvent most of times; van der Waals contacts approach their optimal values, and so on. The presented scenario should unite all contributions which provide the marginal protein stability $\Delta G$. For a typical globular protein $\Delta G$ ranges from -5 to -10Kcal/mol [12].

We set up a system in which, from a thermodynamic point of view, only solvent-chain interactions are taken into account, and are mainly ruled by the hydrophobic effect. It allow us handle with an entropic formulation for the folding problem. We assume that in a composite system, constituted by a protein–like chain surrounded by its solvent molecules, the occurrence chance of a specific chain configuration is expressed as a function of the constraints that it imposes on the rest of the system, the solvent [13]. This approach is applied in a minimalist lattice model, a non-native centric one, managed by Monte Carlo (MC) simulations. The model incorporates a specific additive potential that reproduces just the hydrophobic effect. For each chosen "protein", a particular target, one suitable sequence of "residues" is designed. Our approach has to do with the inverse protein folding problem [10, 11], and it is different from those involving a diffusive search for the state of minimum potential energy. Further, we elaborated a specific sampling method to control the chain configurational evolution. We use the evidence that near room temperature the hydrophobic effect has a remarkable entropic nature. This fact enables us to exploit the transfer free energy to specify the transition probability between two consecutive chain configurations. Our sampling method also includes the effect of local thermal fluctuations which has a significant impact over the efficiency on the search mechanism, because the typical range of proteins is the nanoscale. The proposed sampling was achieved with aid of a central expression coming from Nonextensive Statistical Mechanics (NSM), the generalized $q$-exponential. NSM is a proposed generalization for the usual Boltzmann-Gibbs Statistical Mechanics [14].

The results presented in this work were accomplished through a massive number of MC simulations of 200 representative targets. Besides in-depth support for some results already pointed out in our previous works [10, 11, 13, 15-17], one of our new results shows the existence of a strong correlation between the hydrophobic component of the protein stability and its folding rate.

## 2 – The hydrophobic effect as the main driving-force for folding.

Along the first stage of the folding process we assume that intrapeptide HBs are negligible in order to focus only on the hydrophobic effect. So we consider an isolated heterogeneous system with large and constant volume $V_0$, energy $E_0$, $N_0$ solvent molecules, plus just one protein chain. Once the chain-solvent system is closed, it is supposed has entropy $S_0 = k_B \ln(\Gamma_0)$, here $k_B$ is the Boltzmann constant and $\Gamma_0$ is the number of all possible equiprobable configurations of the composite system, each one has the occurrence probability $p = 1/\Gamma_0$. Due to the large size of the whole system, the temperature $T$, as well as any other intensive parameter, is not affected by whatever the chain could do. Given such specifications one can imagine a thought experiment in which the statistical essence of the solvent-chain interplay can be captured: Suppose that the chain, here seen as a subsystem of the whole system [18-20], has its spatial coordinates fixed in a specific configuration $C$. If, a number of $\Gamma_C$ configurations ($\Gamma_C < \Gamma_0$) is left to the medium, then the entropy is a function of $C$, that is, $S_C = k_B \ln(\Gamma_C)$. It is immediate to see that distinct chain configurations can limit the number of solvent configurations by different amounts, so that the chain configurations which will turn out more times are those who make $\Gamma_C$ as great as possible. Hence, the occurrence probability $P_C$ of any particular chain configuration $C$ is given by $P_C = \Gamma_C/\Gamma_0$. So the larger is $\Gamma_C$, greater is the occurrence chance of configuration $C$. In other words, the occurrence chance of a given configuration $C$ depends on how much restrictions the chain imposes to the medium. In terms of the entropy change we have

$$P_C = \exp(\Delta S_C/k_B). \tag{1}$$

***Transition probability.*** For the sake of computational time, the whole system, chain plus solvent, is assumed to be in the macroscopic equilibrium, even when the chain is far away from its native conformation along the search stage. Therefore, we conjecture that the detailed balance is satisfied. Anyway it is known that MC trial moves can drive a system from a peculiar non-equilibrium scenario to the equilibrium condition [20, 21]. Thus, the transition probability $T_{a \to b}$ from configuration $a$ to configuration $b$ is set as $T_{a \to b} = P_a/P_b$, when $P_a < P_b$. So Eq.1 can be written as

$$T_{a \to b} = \exp(\Delta S_{ab}/k_B), \tag{2}$$

here $\Delta S_{ab} = S_b - S_a$ is the difference between the entropies of the whole system, when the chain moves from configuration $a$ to configuration $b$. Once the system is closed, the microcanonical ensemble condition, the energy changes $\Delta E_{ab} = 0$, then the configurational evolution of the system is only driven by the entropy. Therefore, follows that $\Delta S_{ab} = \Delta G_{ab}/T$. However, it turns out that at room temperature, the condition of almost null enthalpy change is also verified by apolar molecules in the canonical ensemble context by experiments [22], as discussed in the following paragraph.

The hydrophobic effect, seen as the tendency to reverses the limited mobility of water molecules imposed by the solvation shell of the non-polar solutes is a driving force in the sense of $\Delta G < 0$. From the thermodynamic point of view, the change $\Delta G$ can be identified as the transfer free energy, that is, the energy variation to transfer a molecule from one solvent environment to another. Here it means the change of free energy when apolar solutes are transferred from an aqueous to an apolar environment [23]. Data coming from experiments performed with apolar molecules show that enthalpy $\Delta H$ and entropy $T\Delta S$ components of the transfer free energy, $\Delta G = \Delta H - T\Delta S$ increase monotonically within the interval from near zero up to 100°C, and $\Delta G$ almost remains the same. However, about room temperature $\Delta H/T\Delta S \simeq 0$ [22], thus $\Delta H$ can be neglected. One can see this last point at Table 3-1, p. 85 of reference [5]. Therefore, under such condition, the hydrophobic effect can be characterized as an entropic-driven force, and so we write Eq.2 as

$$T_{a \to b} = \exp\left[(-\Delta G_{ab}^h)/k_B T\right], \tag{3}$$

here $\Delta G_{ab}^h$ is the transfer free energy change of the whole system when the chain undergoes a transition between two successive configurations $a \to b$. In its complete form Eq.3 becomes

$$T_{a \to b} = \min\left\{1, \exp\left(-\frac{\Delta G_{ba}^h}{k_B T}\right)\right\}, \tag{4}$$

here $\min\{x, y\} = x$ if $x < y$, and it is equals to $y$ otherwise; the minimum value for the pair $(x, y)$.

***Local thermal fluctuations.*** As the Brownian motion suggests, at nanoscale domain, the thermal noise plays an important role over the conformational behavior of macromolecules. In order to take the thermal noise into account, instead of apply the usual Boltzmann weight, we adopt Tsallis weight to sample chain configurations in our MC simulations [14, 15]. Thus Eq.4, becomes

$$T_{a \to b} = \min\left\{1, \left[1 - (1 - q)\Delta G_{ba}^h / k_B T\right]^{1/(1-q)}\right\}, \tag{5}$$

here we replaced the conventional exponential by its generalization, the $q$-exponential:

$\exp_q(-x) = \left[1 - (1 - q)x\right]^{\frac{1}{1-q}}$, one should note that when $q \to 1$ the usual exponential is recovered. In our approach, the entropic index $q$ is regarded as a dynamical variable, $q$ is associated with the instantaneous degrees of freedom $n$ of an evolving chain by the relation $q = 1 + 2/n$ [24, 25]. In the present approach, $q$ and $n$ are related with each particular configuration of the chain. This is evaluated by the instantaneous compaction of the chain, so

$$q = 1 + 2(q_{max} - 1)/R_G^2, \tag{6}$$

here $R_G$ is the instantaneous radius of gyration of the chain, and $q_{max} = 4/3$, for more details see reference [15]. Given the instantaneous spatial position of the molecule, the entropic index $q$ reflects the effect of local thermal fluctuations acting on the chain. The effect is more pronounced over compacted configurations, the smaller $R_G$ situations, than over those extended ones, where $q \to 1$. For comparison, the effect of Tsallis weight over the transition probability Eq.5, is similar to that of Boltzmann weight when the reservoir temperature $T$ in the argument of the second weight is increased by some amount $\Delta T$. The increase in $T$ depends on each specific target; it has to do with topological features of this specific target. In other words, when one uses the Boltzmann weight, a quite similar sampling can be achieved by increasing the original reservoir temperature $T$, however the specific increment depends on the particular chosen target [15]. Surely this is not what is desired, because for all proteins of the same organism, the folding process must work fine at the same temperature, or at least at the same narrow temperature interval [15]. The net effect of variable $q$ over the chain's configurational evolution is overcome energetic/steric traps easiest than with the standard Boltzmann weight. We give details of such claim in the following topic.

## 3 – The model

The concepts presented in the previous section are applied in the stereochemical model, a minimalist, non-native-centric lattice model consisting of several 27-mer chains embedded in a cubic lattice. Beads represent "residues", they occupy consecutive and distinct sites in a 3D infinite cubic lattice. The remainder lattice sites are filled by "solvent molecules" [11, 16]. Solvent and chain beads interact with each other; they exchange their respective sites along the MC simulation in order to reproduce the behavior of a single protein-like chain in solution. Native structures are represented by compact self-avoiding (CSA) configurations; they are 3×3×3 cubes [26]. Each protein-like chain is designed according to a chosen target. This is done with aid of a reduced repertory composed by a 10-letter alphabet of distinct hydrophobic levels $\{h_k\}$, and a set of inter-residue steric specificities $\{c_{i,j}\}$. The set $\{c_{i,j}\}$ is fixed for each monomer pairs in contact, that is, it does not depend on any particularities of the native structure. These stereo restrictions emulate distinct shapes and sizes features of real residues by establishing what residues pairs are allowed to be lattice first neighbors through hard-core interactions [11]. For each target, the corresponding sequence is elaborated by a design syntax, which is based on the "hydrophobic inside rule", the positive design [27], and also on steric attributes of residues, the negative design [11, 16].

For lattice models, the solvent-chain interaction can be reproduced in an exact way when an additive potential to the first-neighbors is applied, that is, $g_{i,j} = h_i + h_j$, here $h_i$ is the hydrophobic level of the $i$-th residue along the chain sequence [13]. Therefore, from the energetic point of view, each configuration $C$ is characterized by the amount $G_C = \sum_{(i,j)} (g_{i,j} + c_{i,j}) \Delta_{i,j}$. The sum runs over all $(i,j)$ residue pairs: $\Delta_{i,j} = 1$ if the pair $(i,j)$ makes a first neighbor topological contact, and is $\Delta_{i,j} = 0$ otherwise. Changes from configuration $a \rightarrow b$ are accepted, or rejected according to the variation $\Delta G_{a,b}$, see Eq. 4. As described in our previous works [10, 11, 15-17], the present model encompasses some of basic characteristics of the protein folding problem. And also discussed before, the effect of local thermal fluctuations on the configurational evolution is taken into account when one inserts Eq. 5 in the sampling algorithm; this leads to a modified Metropolis $q$ algorithm ($M_q A$).

## 4 – Results

In order to attest the effect of local thermal fluctuations over the folding kinetics, first we performed thousands independent MC simulations using the standard Metropolis Algorithm (MA) and the Metropolis $q$ algorithm ($M_qA$), in which we employ Boltzmann and Tsallis weights, respectively. For each statistical weight, we obtained a set of $10^3$ independent folding times ($\{t_i\}$) for the target labeled 866, a particular structure from a catalogue of 51704 distinct targets. All simulations started from a full random opened chain, that is, a chain configuration without any kind of topological contact, and they finish when the corresponding target structure is found for the first time within a fixed time window ($t_w = 10^5$ MC-time). We identify the spent time to reach the target with the first passage time (fpt). Here, one unit of MC-time (MCt) corresponds to 8100 MC steps, that is, 8100 single chain move attempts.

**Comparison between MA and $M_qA$ algorithms.** Fig.1 shows the comparison between two kinetic curves: the data coming from $M_qA$ is well fitted by just one single exponential (red curve), there we estimate folding characteristic time as $\tau_q = 28$ (MCt). On the other hand, the standard MA predicts $\tau = 80$ (MCt), which is almost three times greater than the previous one, (inset, black curve). In the second case, the best fitting curve was achieved by three decaying exponentials. The obtained result with the standard MA suggests that alternative slow routes to the native were visited, leading the chain to get stuck in wrong conformations for long time intervals. On the contrary, when local thermal fluctuations are taken into account, by means of $M_qA$, wrong packed chain are readily dismounted avoiding energetic/steric traps which imply long lasting paths to the native structure, even for fast targets.

The ratio $\tau/\tau_q \simeq 1/3$ expresses by itself the $M_qA$ versatility. The structure 866, one of the fastest targets, was chosen for the sake of CPU time in order to compare MA and $M_qA$ algorithms. With both algorithms, MA and $M_qA$, all natives for the 866 target were found, without exceptions, within the pre-established time window $t_w$. However, different from the $M_qA$, the standard MA does not work for all protein-like sequences which can be designed for all 51704 distinct targets of the stereochemical model. The situation is, for many sluggish proteins, their natives can only be found, within a reasonable time window, if the $M_qA$ is employed. So, the $M_qA$ amplifies substantially the chance of the chain get rid of energetic and stereo traps, which in some cases can house misfolded conformations virtually forever. The reason behind the $M_qA$ efficiency seems to be due to the fact that, as more compacted the chain configuration is, more significant is the effect of local thermal fluctuations over the globule. Hence, as a matter of explanation, similar to the action of chaperones in larger proteins, when the chain is caught in some wrong compacted conformation, the globule could be more easily dismantled when local thermal fluctuations act upon it, so the folding process can be restarted. We call attention to the fact that some classes of proteins, the small ones, do not need any help from other biological structures to reach their native structures [2].

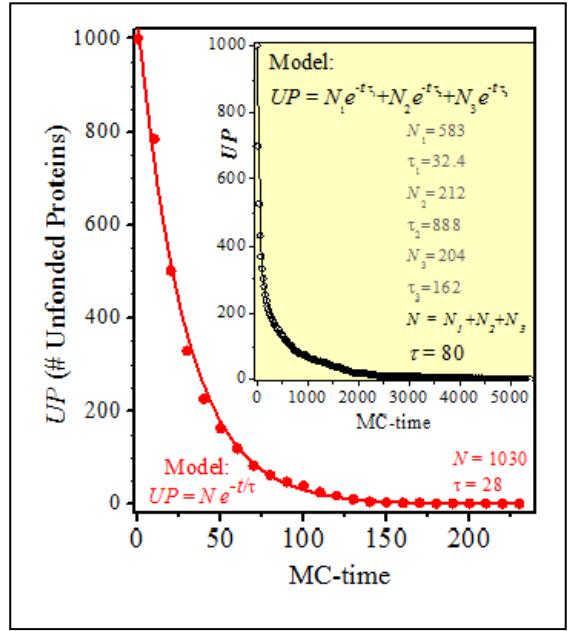

Figure 1 – Kinetic curves for the decay of unfolded proteins (*UP*). Serial independent MC simulations represent the folding process of a diluted solution populated with $10^3$ identical unfolded proteins. **Red curve**, $M_qA$: in order to find the native the exponential behavior of the evolution number of the *UP* suggests a stochastic process with probability $p = 1/\tau_q$ ; $\tau_q = 28$ ( Local thermal fluctuations unpack wrong compacted conformations in a natural and efficient way. Here the entropic index *q* is a variable which depends on the compactness of the globule. **Black curve**, standard MA: when we apply Boltzmann weight the folding process may follow alternative folding routes with some lag; here this condition is described by three decaying exponentials of different characteristic times: $\tau \approx 30$; 160 and 890 (MCt). Details for data treatment to reproduce the ensemble behavior of the folding process can be found in our previous works [10, 11, 28, 29].

***The search mechanism.*** Next, we tested the reliability for the proposed search mechanism in a very sharp and also exhaustive way: We selected a set of 200 targets with diversified configurational complexity. An appropriated sequence of residues was designed for each target [11, 16] and then submitted to 600 independent MC simulations, the $M_qA$ was adopted. For each target *j* ( *j* = 1, 2, … 200) we obtained a set of 600 folding times $\{t_i\}^j$ (*i* = 1, 2, … 600). All $1.2 \times 10^5$ simulations were successful, without any exception, that is, the corresponding natives were always found within the *a priori* established time window $t_w$. This result, by itself, is a reasonable indication of the consistency of the proposed search mechanism for the first stage of the folding process, which assumes that the mechanism is driven only by hydrophobic forces. A key property of hydrophobic effect can explain the success of the search stage: the hydrophobic interactions are not specific, that is, hydrophobic parts of the chain tend to get together but the real interactions are between apolar parts of the protein and solvent molecules. Therefore, besides the hydrophobic interactions be important restrictors to the conformational space, local residues positions can be easily exchanged, so their reorganization along the search stage becomes simpler.

The set $\{t_i\}^j$ is used to determine the folding characteristic time $\tau^j$ for each target "$j$". This was done by means of a similar procedure to the experimental determination of the folding rate $k_f$ ($k_f = 1/\tau$) for real proteins at low concentration; see details in [11,16]. For some targets the folding is very fast, but for others it is too slow, indicating the effect of the native configurational complexity over the process. Several elementary structural patterns influence $\tau$: some of them, when properly combined, allow a very quick folding, while some others limit the folding speed. We show in Fig.2A how $\tau_q$ is limited by increasing the number of certain structural patterns that slow down the folding process, like that one depicted in the inset. On the other hand, in part B, we show how the folding process can be speed up with the increasing of the combination of structural patterns that resemble with $\alpha$-helices, or $\beta$-strands [16], like the example presented in the inset. The intricate combination of favorable and unfavorable patterns seems to be a key determinant for $\tau_q$. Indeed, despite its simplicity, when the stereochemical model is simulated by M$_q$A it reveals that $\tau_q$ ranges over four orders of magnitude. Therefore, from a qualitative perspective, our results suggest that $\tau_q$ is modulated by the way that specific elementary structural patterns are combined.

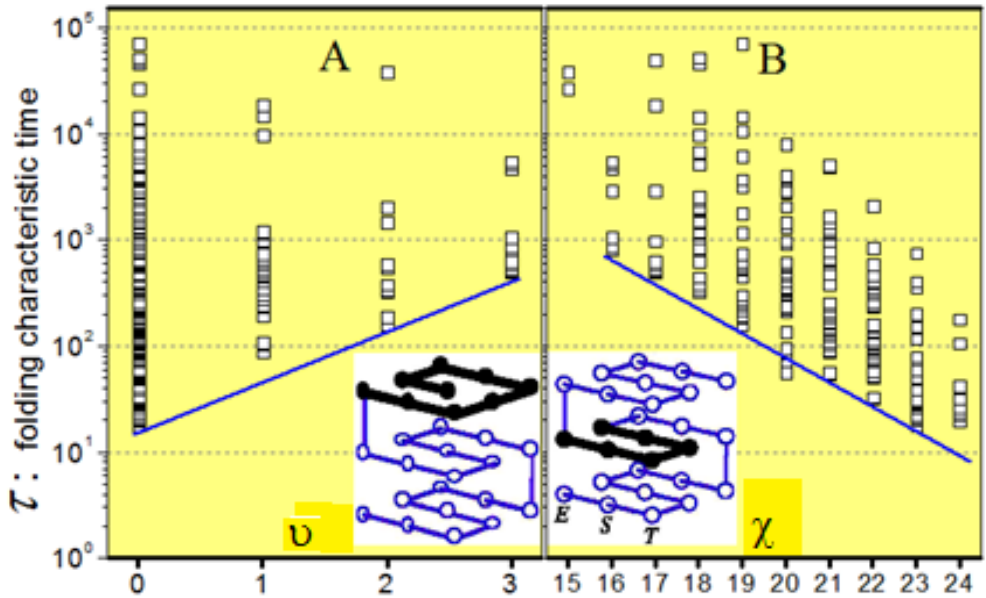

**Figure 2** $\tau$ is a function of the number of combination of certain structural patterns, regardless the statistical weight adopted to perform MC simulations. Once we adopted the $M_q A$, the open squares represent $\tau_q$ for each one of those 200 chosen targets, their values range in the interval $[10 - 10^5]$ and are plotted as a function of the number of two parameters, $\upsilon$ and $\chi$. The $\chi$ parameter is the number of favorable structural patterns that speed up the folding kinetics, while $\upsilon$ is the number of patterns that slow down the kinetics. In part **B** residues are labeled according with its relative environment in the native structure: residues at the chain ends (**E**), residues positioned in turns (**T**) and those residues found in straight lines (**S**). Once this previous definition was settled, one can find some sequences specified by the combination "TSTSTSTTE", like that highlighted in the inset of part **A** that made the kinetic access to the native difficult, in a non-cooperative sense. As the number of $\upsilon$ parameter increases ($\upsilon = 1$, 2 or 3), $\tau_q$ increases too. On the right side, part **B**, when planar sequences like "STTS", are combined in propitious ways, they can speed up the process. So one can find lower $\tau_q$'s. Here the pictured target in **B** has 8 $\chi$ sequence patterns, they are disposed in a cooperative way. So the process is speeded up, this give us a smaller $\tau_q$. As the number $\chi$ increases ($\chi = 1, 2, \ldots, 24$), $\tau_q$ decreases proportionally. We stress that the way these structural patters, $\upsilon$ and $\chi$, are combined determines the folding speed for each chosen target.

***Folding rate and stability.*** For small globular proteins with two-state kinetics, single global parameters as contact order (CO), or absolute CO (ACO) correlate with the constant rate of folding $k_f$ [30, 31]. Despite such correlation be persuasive from the statistical point of view, it lacks of clear physical arguments to explain the overall behavior of two-state kinetics. Many exceptions are always found, with points out of the CO×ln($k_f$) fitted linear regression. The result presented in Fig.2 suggests that the many ways of organize elementary structural patters like helices, strands, or loops in cooperative or non-cooperative patterns can help to explain why it is usual plot CO×ln($k_f$) for proteins grouped according some specific patterns, for instance, all α, α+β, α/β, and so on [16, 30]. On the other hand, folding rates of small two-state proteins have been considered unrelated with the stability of their respectives native states [30]. However, this point remains as an open problem [32]. In this context, stability, the free energy difference between the folded and unfolded states, is evaluated by $\Delta G = -RT\ln(K_{eq})$, here $R$ is the gas constant, $T$ is the absolute temperature, and $K_{eq}$ is the equilibrium constant. For two-state proteins $K_{eq} = [N]/[U]$, here $[N]$ and $[U]$ are the concentration of folded and unfolded proteins.

The equilibrium relation tells us about the balance between [N] and [U]. It doesn't means that all same proteins keep themselves on their native or unfolded states over all the time. At marginal equilibrium and low concentration conditions, one particular protein in the native state eventually unfolds, and returns to the native later on. So, from a series of $M$ snapshots taken over a sufficient long time interval for the same protein, the number $m_N$ out of $M$ snapshots in which the protein figures in the native state, give us a kinetic view of $K_{eq}$. From the equilibrium relation given before we get $K_{eq} = \phi/(1-\phi)$, here $\phi = m_N/M$ is the relative frequency $\phi$ in which the protein figures in the native state. Then $\Delta G$ can be written as

$$\Delta G/RT = -\ln(\phi/(1-\phi)). \tag{7}$$

With the aim to determine $\phi$ we performed a set of extended MC runs for all 200 targets. Each simulation starts with the chain in the native structure: Along the simulation we took a sample of $M$ configurations equally separated in time, and we count the number $m_N$ of configurations that match the native. The time window for the run depends on the target and must be large enough to produce a meaningful statistics. For the present purpose, the native conformation was defined as any configuration with at least 24 out of its total 28 native contacts. All 3x3x3 CSA configuration has a total of 28 topological contacts.

As described before, the set $\{\tau^j\}$ shown in Fig.2 and $\{\phi^j\}$ obtained from Eq.7 for all targets ($j = 1,2, \dots, 200$), are used to verify how the folding rate $k_f$ ($k_f = 1/\tau$) relates with the stability $\Delta G$ (Eq. 7), which contains only the hydrophobic component of the overall structural stability. As shown in Fig.3, $\Delta G^h/RT$ correlates expressively with the logarithm of the folding rate $\ln(k_f)$, the Pearson coefficient is $R = 0.85$.

The found correlation is remarkable. Since Kauzmann seminal work [4], it became ordinary to point to the hydrophobic effect as the main factor for protein stability. However, contrasting with this view, the result shown in Fig.3 suggests that the hydrophobic stability for a specific target depends, in a very sharp way, with its folding rate. As discussed before, $\tau$ also depends on the native structural complexity. Therefore, the hydrophobic effect could not be a general stabilizing factor for proteins. On the other hand, an opposite point of view has claimed that the interpeptides HBs provide the main driving force for folding, at the same time that they assure the protein stability [8, 9, 33]. However, the protein folding process seems to be more subtle: The present results give us the insight that the native stability could depend on a number of assorted contributions; like hydrophobic reactions and intra-chain interplay, as backbone HB and Van der Waals interactions, the amount of each contribution would also depend on the complexity of the native structure.

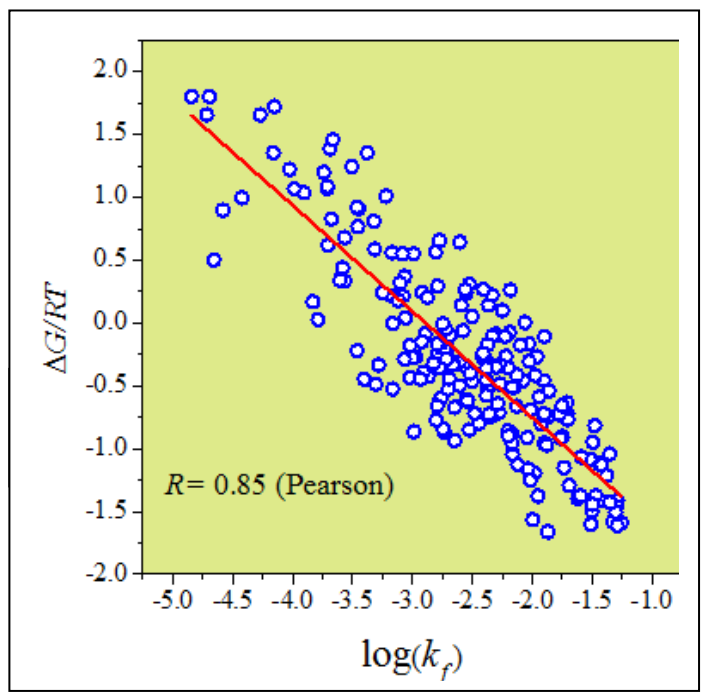

Figure 3 - The hydrophobic stability ($\Delta G/RT$) correlates with the folding rate $k_f = 1/\tau_q$. It's noteworthy that the simple model applied here predicts $k_f$ that covers four orders of magnitude. This finding reflects the influence of the target's structural complexity. The free energy values are normalized; here the zero value is located around the $\Delta G/RT$-axis center.

## 5 – Comments and conclusions

This manuscript is mainly concerned to the first stage of the folding process, the search mechanism. We assumed as a working hypothesis that the first stage is an autonomous kinetic process ruled only by the hydrophobic effect. Although our minimalist model does not comprise other stabilizing forces present in the second step, our results suggest that the hydrophobic component of protein stability is modulated by the topological complexity of the native.

The lattice model employed here emulates a protein in aqueous solution: a chain constituted of 27 beads is immersed in an infinite cubic lattice, in which the remainder sites represent the solvent. For each chosen target, a 3×3×3 cube, a protein-like chain was designed from a 10–letter alphabet, whose constituents are discriminated by their hydrophobic levels and also steric attributes. Only solvent-chain and hardcore intrachain interactions are considered. This sterochemical model is able to reproduce some general characteristics of the protein folding problem; for details see references [10, 11, 15, 16].

Characteristic folding times, $\tau$ and $\tau_q$, estimated by means of two sampling techniques, the standard MA and $M_qA$, demonstrate the superior performance of the latter one, reflected by the ratio $\tau/\tau_q \simeq 1/3$; Fig.1. The importance of the $M_qA$ method becomes more dramatic for those sluggish proteins. For many of them the MA method never finds the target within a reasonable finite time window. The key aspect of the Tsallis weight, as employed here, is that the entropic index $q$ becomes a variable which depends on the instantaneous compactness of the chain; see Eq.5 and reference [15]. As more compacted is the chain, larger are local thermal fluctuations acting on the globule, increasing the probability to get rid of wrong compacted conformations. Similar to the random movement observed in the Brownian movement [34], a compacted chain conformation experiences more significant fluctuations due to the microscopic chaos than an extended chain conformation.

It is remarkable that some single-domain proteins with similar size and stability, like *Escherichia coli cytochrome b₅₆₂* (pdb id: 1qpu; 106 residues) and muscle *Acylphosphatase* (pdb id: 2acy; 98 residues), have folding times differing by almost three orders of magnitude. Several protein folding theories have been invoked to explain this experimental fact [35]; amongst them the hypothesis of topology-dependent rates has gained much attention [30, 31, 35]. By its own turn, our outcome about the characteristic folding times of 200 representative structures of assorted structural complexities, also corroborates the premise that the overall structural topology is a strong determinant of folding rates: the elements of the set $\{\tau_q\}$ cover four orders of magnitude, and their values are limited by the structural composition of structural patterns. As the number of the unfavorable structural pattern $\upsilon$ increases in the native, their values for $\tau_q$ also increases; inset of Fig. 2A. On the other hand, when the number of favorable structural patterns $\chi$ increases, $\tau_q$ is reduced; inset of Fig. 2B. Further details about $\upsilon$ and $\chi$ are given in reference [16]. In addition, Fig.2 suggests that different subtle combinations of the same set of structural patterns can also modulate $\tau_q$ in a substantial way. As a matter of robustness, the extensive set of 600 successful MC simulations for each one of those 200 distinct targets, which sums up more than $10^5$ independent runs, attests the sufficiency and reliability of the hydrophobic effect as the main driving force for the first stage of folding process, while the native topology, composed by combinations of structural patterns, rules the temporal scales for kinetics.

Further the stereochemical model propose that the hydrophobic component of stability, $\Delta G^h$, depends on its folding rate $k_f$ in a strong way; see Fig.3. At first sight, one could argue that such correlation would be due to the eventual interconnection between the "hydrophobic charges" of the target with the folding rate. However, this not seems to be the case, once it has been shown that the "hydrophobic charge" does not correlate with CO: many cases that present fast kinetics have small CO values, but they show less "hydrophobic charges" than many others who have slow kinetics, and high CO values [36]. Therefore, the kinetic accessibility to the native seems to be the central factor in order to explain the obtained correlation $\Delta G^{h} \times k_{f}$. In other words, proteins which get their natives so quick, when leave them, they can come back so fast, and *vice-versa*.

We stress that, at the best of our present knowledge, there is not any obvious, or even clear reason for the scope of the correlation showed here $\Delta G^{h} \times k_{f}$ not be broader. Therefore, despite of our results diverge from the usual notion that the protein stability should be originated from hydrophobic interactions between non polar groups [5], they gave us the insight that the native stability seems to depend on a number of assorted contributions. They could came from the interplay between hydrophobic reactions and intra-chain, like backbone HB and Van der Waals interactions, whose amount of each contribution also may depend on the geometrical complexity of the native structure.

## ACKNOWLEDGMENT
We thank PNPD/CAPES for funding, as well as we thank referees for their questions and also valuable suggestions.

## REFERENCES

[1]- R.H. Pain, editor, Mechanisms of Protein Folding, 2$^{nd}$ edition, Oxford University Press, Oxford, 2000.

[2]- C.B. Anfinsen, Science. 181 (1973) 223.

[3]- P.G. Wolynes, Biochimie. 119 (2015) 218.

[4]- W. Kauzmann, Advances in Protein Chemistry. 14 (1959) 1.

[5]- P.L. Privalov, in Protein Folding, edited by T.E. Creighton, W.H. Freeman, New York, 1992.

[6]- K.A. Dill, S.B. Ozkan, M.S. Shell, and T.R. Weikl, Annu. Rev. Biophys. 37 (2008) 289.

[7]- A.V. Morozov, T. Kortemme, Adv. Protein. Chem. 72 (2005) 1.

[8]- G.D. Rose, P.J. Fleming, J.R. Banavar, and A. Maritan, Proc. Natl. Acad. Sci. USA 103 (2006) 16623.

[9]- A. Ben-Naim, The Protein Folding Problem and its Solution, Word Scientific Publishing Company, Singapore, 2013.

[10]- M.E.P. Tarragó, L.F.O. Rocha, R.A. da Silva, and A. Caliri, Phys. Rev. E. 67 (2003) 031901.

[11]- R.A. da Silva, M.A.A. da Silva, and A. Caliri, J. Chem. Phys. 114 (2001) 4235.

[12]- E.E. Lattman and G.D. Rose, Proc. Natl. Acad. Sci. USA. 90 (1993) 439.

[13]- L.F.O. Rocha, I.R. Silva, and A. Caliri, Physica A. 388 (2009) 4097.

[14]-    C. Tsallis, Introduction to Nonextensive Statistical Mechanics - Approaching a Complex World, Springer, NewYork, 2009.

[15]-    J.P. Dal Molin, M.A.A. da Silva, and A. Caliri, Phys. Rev. E. 84 (2011) 041903.

[16]-    I.R. Silva, L.M. Dos Reis, and A. Caliri, J. Chem. Phys. 123 (2005) 154906.

[17]-    L.F.O. Rocha, M.E.P. Tarragó, and A. Caliri. Braz. J. Phys. 34 (2004) 90.

[18]-    L.D. Landau and E.M. Lifshitz, Statistical Physics, Part 1: Volume 5 - Course of Theoretical Physics, 3rd Edition, Butterworth-Heinemann - Elsevier, China, reprint 2011.

[19]    D.L. Goodstein, States of Matter, Prentice-Hall, New York, 1975.

[20]-    N. Metropolis, A. Rosenbluth, M. Rosenbluth, A. Teller, and E. Teller, J. Chem. Phys. 21(1953) 1087.

[21]-    T.R. Gingrich and P.L. Geissler, J. Chem. Phys. 142 (2015) 234104.

[22]-    N.T. Southall, K.A. Dill, and A.D. Haymet, J. J. Phys. Chem. B. 106 (2002) 521.

[23]-    E.A. Mills and S.S. Plotkin, J. Phys. Chem. B. 117 (2013) 13278.

[24]-    G. Wilk and Z. Wlodarczyk, Phys. Rev. Lett. 84 (2000) 2770.

[25]-    C. Beck, G.S. Lewis, and H.L. Swinney, Phys. Rev. E. 63 (2001) 035303(R).

[26]-    E. Shakhnovich, G. Farztdinov, A.M. Gutin, and M. Karplus,  Phys. Rev. Lett. 67 (1991) 1665.

[27]-     K.A. Dill et al. Principles of protein folding -A perspective from simple exact models.  Protein Science, 4 (1995) 561.

[28]-    J.P. Dal Molin, and A. Caliri, Current Physical Chemistry. 3 (2013) 69.

[29]-    J.P. Dal Molin, M.A. Alves da Silva, I.R. da Silva, and A. Caliri, Braz. J. Phys. 39 (2009) 435.

[30]-    K.W. Plaxco, K.T. Simons, and D. Baker, J. Mol. Biol. 277 (1998) 985.

[31]-    M. Rustad and K. Ghosh, J. Chem. Phys. 137 (2012) 205104.

[ 32]-    A.R. Dinner and M. Karplus, Nat. Struct. Biol. 8 (2001) 21.

[33]-    A. Ben-Naim, Open Journal of Biophysics. 1 (2011) 1.

[34]-    P. Gaspard, M.E. Briggs, M.K. Francis, J.V. Sengers, R.W. Gammon, J.R. Dorfman and R.V. Calabrese, Nature. 394 (1998) 865.

[35]-    B. Gillespie and K.W. Plaxco, Using protein folding rates to test protein folding theories,  Annu. Rev. Biochem. 73 (2004) 837.

[36]-    I.R. Silva. Enovelamento proteico: fatores topológicos. Doctoral Thesis, Universidade de São Paulo, 2005, pg 45, in Portuguese: http://www.teses.usp.br/teses/disponiveis/59/59135/tde-24082005-112844/pt-br.php